\documentclass[11pt]{article}
\usepackage{amsmath,amsfonts,amssymb,graphicx,epsfig,pdflscape}
\usepackage[usenames]{color}
\usepackage{pstricks}
\numberwithin{equation}{section}

\newcommand{\nc}{\newcommand}
\nc\disp{\displaystyle}
\nc{\fh}{\hat{f}}
\nc{\muh}{\hat{\mu}}
\nc{\nuh}{\hat{\nu}}
\nc{\spos}[2]{\makebox(0,0)[#1]{$\sm{#2}$}}
\nc{\sm}[1]{{\scriptstyle #1}}
\nc{\al}{\alpha}
\nc{\g}{\gamma}
\nc{\G}{\Gamma}
\nc{\D}{\Delta}
\nc{\eps}{\epsilon}
\nc{\la}{\lambda}
\nc{\La}{\Lambda}
\nc{\var}{\varphi}
\nc{\pa}{\partial}
\nc{\nn}{\nonumber \\ }
\nc{\hf}{\frac{1}{2}}
\nc{\dz}{\frac{dz}{2\pi i}}
\nc{\bin}[2]{\left(\!\!\!\begin{array}{c} {#1}\\ {#2} \end{array}\!\!\!\right)}
\nc{\bea}{\begin{eqnarray}}
\nc{\eea}{\end{eqnarray}}
\nc{\bra}[1]{\langle {#1}|}
\nc{\ket}[1]{|{#1}\rangle}
\nc{\ketw}[1]{({#1})_{{\cal W}}}
\nc{\chit}{\raisebox{0.25ex}{$\chi$}}
\nc{\chih}{\raisebox{0.25ex}{$\hat\chi$}}
\nc{\Db}{\mbox{\boldmath $D$}}
\nc{\Hb}{\mbox{\boldmath $H$}}
\nc{\calH}{{\cal H}}
\nc{\calR}{{\cal R}}
\nc{\calL}{{\cal L}}
\nc{\calV}{{\cal V}}
\nc{\Hc}{{\cal H}}
\nc{\Rc}{{\cal R}}
\nc{\Lc}{{\cal L}}
\nc{\Vc}{{\cal V}}
\nc{\Ib}{\mbox{\boldmath $I$}}
\nc{\qb}{\bar{q}}
\def\vvdots{\mathinner{\mkern1mu\raise1pt\vbox{\kern7pt\hbox{.}}\mkern2mu
  \raise4pt\hbox{.}\mkern2mu\raise7pt\hbox{.}\mkern1mu}}
\nc{\gauss}[2]{\left[\!\!\begin{array}{c} {#1}\\ {#2} \end{array}\!\!\right]}
\nc{\sbin}[2]{\left\{\!\!\!\begin{array}{c} {#1}\\ {#2} 
\end{array}\!\!\!\right\}}
\nc{\sbinlr}[2]{\Big\langle\!\!\begin{array}{c} {#1}\\ {#2} 
\end{array}\!\!\Big\rangle}
\nc{\bino}[2]{\left(\!\!\begin{array}{c} {#1}\\ {#2} \end{array}\!\!\right)}
\def\half {\mbox{$\textstyle \frac{1}{2}$}}

\definecolor{lightblue}{rgb}{.61,.61,1}
\definecolor{midblue}{rgb}{.7,.7,1}
\definecolor{lightlightblue}{rgb}{.85,.85,1}
\definecolor{lightestblue}{rgb}{.96,.96,1}
\definecolor{lightpurple}{rgb}{1,.65,1}

\def\loopa{
\psframe[linewidth=.25pt](0,0)(1,1)
\psarc[linewidth=1.5pt,linecolor=blue](1,0){.5}{90}{180}
\psarc[linewidth=1.5pt,linecolor=blue](0,1){.5}{-90}{0}
}
\def\loopb{
\psframe[linewidth=.25pt](0,0)(1,1)
\psarc[linewidth=1.5pt,linecolor=blue](0,0){.5}{0}{90}
\psarc[linewidth=1.5pt,linecolor=blue](1,1){.5}{180}{270}
}

\def\facegrid#1#2{
\psframe[fillstyle=solid,fillcolor=lightlightblue,linewidth=0pt]#1#2
\psgrid[gridlabels=0pt,subgriddiv=1]#1#2}
\def\facegridblue#1#2{
\psframe[fillstyle=solid,fillcolor=lightblue,linewidth=0pt]#1#2
\psgrid[gridlabels=0pt,subgriddiv=1]#1#2}
\nc{\ch}{{\rm ch}}
\nc{\R}{{\cal R}}
\nc{\dkk}{\delta_{j,\{k,k'\}}^{(2)}}
\nc{\drr}{\delta_{j,\{r,r'\}}^{(2)}}
\nc{\ddkk}{\delta_{j,\{k,k'\}}^{(4)}}
\nc{\dddkk}{\delta_{j,\{k,k'\}}^{(8)}}
\nc{\dnn}{\delta_{j,\{n,n'\}}^{(2)}}
\nc{\ddnn}{\delta_{j,\{n,n'\}}^{(4)}}
\nc{\dddnn}{\delta_{j,\{n,n'\}}^{(8)}}

\begin{document}

\topmargin -5mm
\oddsidemargin 5mm

\setcounter{page}{1}

\vspace{8mm}
\begin{center}
{\Large {\bf\protect\boldmath Integrable Boundary Conditions and ${\cal W}$-Extended Fusion\\[8pt]
in the Logarithmic Minimal Models ${\cal LM}(1,p)$}}

\vspace{10mm}
 {\Large Paul A. Pearce\footnote{Email: P.Pearce@ms.unimelb.edu.au}},
 \ \  {\Large J{\o}rgen Rasmussen\footnote{Email: J.Rasmussen@ms.unimelb.edu.au}}
\\[.3cm]
 {\em Department of Mathematics and Statistics, University of Melbourne}\\
 {\em Parkville, Victoria 3010, Australia}
\\[.4cm]
 {\Large Philippe Ruelle\footnote{Email: philippe.ruelle@uclouvain.be}}
\\[.3cm]
 {\em Institut de Physique Th\'eorique, Universit\'e catholique de Louvain}\\
 {\em 1348 Louvain-la-Neuve, Belgium}\\[.4cm]

\end{center}

\vspace{8mm}
\centerline{{\bf{Abstract}}}
\vskip.4cm
\noindent
We consider the logarithmic minimal models ${\cal LM}(1,p)$ as `rational' logarithmic 
conformal field theories with  \mbox{extended} ${\cal W}$ symmetry.
To make contact with the extended picture starting from the lattice, we identify $4p-2$ boundary conditions as specific limits of integrable boundary conditions of the underlying Yang-Baxter integrable lattice models. Specifically, we identify $2p$ integrable boundary conditions to match the $2p$ known irreducible ${\cal W}$-representations. 
These $2p$ extended representations naturally decompose into infinite sums of the 
irreducible Virasoro representations $(r,s)$.
A further $2p-2$ reducible yet indecomposable ${\cal W}$-representations of rank 2
are generated by fusion and these decompose as infinite sums of indecomposable rank-2
Virasoro representations. The fusion rules in the extended picture are deduced from the known fusion 
rules for the Virasoro representations of ${\cal LM}(1,p)$ and are found to be in agreement with 
previous works. 
The closure of the fusion algebra on a finite number of representations in the extended picture
is remarkable confirmation of the consistency of the lattice approach.

\renewcommand{\thefootnote}{\arabic{footnote}}
\setcounter{footnote}{0}

\section{Introduction}

Logarithmic conformal field theories~\cite{Gurarie93,RS92} are vital in studying  the critical behaviour of polymers, percolation and other non-local processes, such as the Abelian sandpile model, that possess a countably infinite number of scaling fields~\cite{Saleur87,Duplantier86,DuplantierSaleur87,Kausch95,Cardy99,GuLu99,JPR06,PGPR08}. 
The properties of {\em logarithmic} theories, however, are very different to the familiar properties 
of {\em rational} conformal field theories. Logarithmic theories are neither rational in the strict sense nor unitary and typically exhibit logarithmic branch cuts in correlation functions. But perhaps the most characteristic property is that they admit reducible yet indecomposable representations of the underlying conformal algebra (Virasoro or one of its extensions).
Some useful reviews on logarithmic conformal field theory can be found in \cite{Flohr03,Gaberdiel03,Kawai03}.

The most studied logarithmic theories to date have central charges
\begin{equation}
 c=c_{p,p'}=1-6\,\frac{(p-p')^2}{pp'},\qquad\mbox{$p, p'\in\mathbb{N}$ coprime}
\label{centralcharge}
\end{equation}
These theories are often referred to as augmented $c_{p,p'}$ minimal models within the 
algebraic approach to logarithmic theories. This algebraic approach has proved to be very 
powerful and has produced a substantial body of results~\cite{Kausch91,Flohr96,GK9604,GK9606,Kausch00,FGST0504,FGST0512,GR06,GR07,MR0708,GTipunin07,MR0711}. 
However, it is convenient for us to work here in a lattice approach and consider the logarithmic minimal models ${\cal LM}(p,p')$~\cite{PRZ}. The logarithmic minimal models have precisely the central charges (\ref{centralcharge}) and are defined in an unambiguous and physically consistent manner through the continuum scaling limit of Yang-Baxter integrable models on the square lattice. 
There is now accumulating evidence from conformal data, the structure of indecomposable representations  and fusion rules~\cite{EF06,RP0706,RP0707} to indicate that the logarithmic 
minimal models ${\cal LM}(p,p')$ actually coincide with the augmented $c_{p,p'}$ minimal 
models.

A very interesting open question concerns the existence of some form of extended conformal 
symmetry in these logarithmic theories. From the algebraic approach, 
it is known~\cite{Kausch91,Flohr96,GK9606} that the augmented $c_{1,p}$ minimal models 
indeed possess an extended ${\cal W}$-algebra symmetry. In the ${\cal W}$-extended picture, 
the countably infinite Virasoro representations are reorganized into a finite number of 
${\cal W}$-representations which close among themselves under fusion. 

In this paper, we consider the logarithmic minimal models ${\cal LM}(1,p)$. 
The first member ${\cal LM}(1,2)$ of this series, with central charge $c=-2$, is {\em critical dense polymers}~\cite{PR06} in the Virasoro picture but {\em symplectic fermions}~\cite{Kausch00} in the ${\cal W}$-extended picture. 
Our central result is to demonstrate the compatibility of the Virasoro and ${\cal W}$-extended pictures of the ${\cal LM}(1,p)$ models within the lattice approach. This is achieved by identifying the integrable boundary conditions corresponding to the ${\cal W}$-extended representations and using the known Virasoro fusion rules to establish the ${\cal W}$-extended fusion rules. 
In the process, we also establish  that the logarithmic minimal models provide lattice 
realizations of symplectic fermions and other logarithmic theories with extended conformal symmetry.

The layout of this paper is as follows. 
In Section~2, we recall properties of the various Virasoro representations (Kac, irreducible, indecomposable), their characters, their corresponding integrable boundary conditions and their associated fusion rules. 
In Section~3, we recall the properties of the ${\cal W}$-representations (irreducible, indecomposable) and their extended characters. 
We use fusion of irreducible Virasoro representations to construct  $2p$ integrable boundary conditions as solutions to the boundary Yang-Baxter equation and identify these with the irreducible ${\cal W}$-representations. These $2p$ extended representations naturally decompose into infinite sums of the irreducible Virasoro representations $(r,s)$. Fusion of the irreducible ${\cal W}$-representations produces a further $2p-2$ reducible yet indecomposable ${\cal W}$-representations of rank 2 which decompose as infinite sums of indecomposable rank-2 Virasoro representations. This yields a total of $4p-2$ representations in the ${\cal W}$-extended picture. Finally, we use the known Virasoro fusion rules to deduce the fusion rules of the ${\cal W}$-representations and find that they are in agreement with previous works. Explicit Cayley tables for ${\cal LM}(1,2)$ and ${\cal LM}(1,3)$ are given in Figure~4. 
Throughout, we use the notation $\mathbb{Z}_{n,m}=\mathbb{Z}\cap[n,m]$, with $n,m\in\mathbb{Z}$, 
to denote the set of integers from $n$ to $m$, both included.

\section{Logarithmic Minimal Model ${\cal LM}(1,p)$: Virasoro Picture}

For a given integer $p>1$, the logarithmic minimal model ${\cal LM}(1,p)$ is defined \cite{PRZ} as a Yang-Baxter integrable model on the square lattice. The face operators are defined in the planar Temperley-Lieb algebra~\cite{Jones} by
%
%
\psset{unit=1cm}
\setlength{\unitlength}{1cm}
\begin{equation}
 X(u)\;=\;
 \begin{pspicture}(-0.1,.4)(1.1,1.1)
\facegrid{(0,0)}{(1,1)}
\psarc[linewidth=.5pt](0,0){.15}{0}{90}
\rput(.5,.5){\small $u$}
\end{pspicture}
 \;=\;\
  \frac{\sin(\lambda-u)}{\sin\lambda}\ 
  \begin{pspicture}(-.1,.4)(1,1.1)
    \facegrid{(0,0)}{(1,1)}
    \put(0,0){\loopa}
  \end{pspicture}
 \ \; +\ \frac{\sin u}{\sin\lambda}\ 
  \begin{pspicture}(-.1,.4)(1,1.1)
    \facegrid{(0,0)}{(1,1)}
    \put(0,0){\loopb}
  \end{pspicture}
\end{equation}
%
%
where $u$ is the spectral parameter and $ \lambda\;=\;\frac{(p-1)\pi}{p}$ is the crossing parameter. 
The relations in the diagrammatic algebra ensure that, in addition to the face Boltzmann weights, each closed loop 
is weighted by the nonlocal loop fugacity $ \beta\;=\;2\cos\lambda$. 
A typical configuration is shown in Figure~1.

\begin{figure}[htbp]
\psset{unit=.8cm}
\setlength{\unitlength}{.8cm}
\bea
\begin{pspicture}(10,7)
\facegrid{(0,0)}{(8,4)}
\facegridblue{(8,0)}{(10,4)}
\psline[linecolor=blue,linewidth=2pt](8.5,0)(8.5,4)
\psline[linecolor=blue,linewidth=2pt](9.5,0)(9.5,4)
\psline[linecolor=blue,linewidth=2pt](8,.5)(8.4,.5)
\psline[linecolor=blue,linewidth=2pt](8.6,.5)(9.4,.5)
\psline[linecolor=blue,linewidth=2pt](9.6,.5)(10,.5)
\psline[linecolor=blue,linewidth=2pt](8,1.5)(8.4,1.5)
\psline[linecolor=blue,linewidth=2pt](8.6,1.5)(9.4,1.5)
\psline[linecolor=blue,linewidth=2pt](9.6,1.5)(10,1.5)
\psline[linecolor=blue,linewidth=2pt](8,2.5)(8.4,2.5)
\psline[linecolor=blue,linewidth=2pt](8.6,2.5)(9.4,2.5)
\psline[linecolor=blue,linewidth=2pt](9.6,2.5)(10,2.5)
\psline[linecolor=blue,linewidth=2pt](8,3.5)(8.4,3.5)
\psline[linecolor=blue,linewidth=2pt](8.6,3.5)(9.4,3.5)
\psline[linecolor=blue,linewidth=2pt](9.6,3.5)(10,3.5)
\rput[bl](0,0){\loopb}
\rput[bl](1,0){\loopb}
\rput[bl](2,0){\loopa}
\rput[bl](3,0){\loopa}
\rput[bl](4,0){\loopa}
\rput[bl](5,0){\loopa}
\rput[bl](6,0){\loopa}
\rput[bl](7,0){\loopa}
\rput[bl](0,1){\loopa}
\rput[bl](1,1){\loopa}
\rput[bl](2,1){\loopa}
\rput[bl](3,1){\loopa}
\rput[bl](4,1){\loopa}
\rput[bl](5,1){\loopa}
\rput[bl](6,1){\loopb}
\rput[bl](7,1){\loopb}
\rput[bl](0,2){\loopb}
\rput[bl](1,2){\loopb}
\rput[bl](2,2){\loopa}
\rput[bl](3,2){\loopb}
\rput[bl](4,2){\loopb}
\rput[bl](5,2){\loopb}
\rput[bl](6,2){\loopb}
\rput[bl](7,2){\loopa}
\rput[bl](0,3){\loopa}
\rput[bl](1,3){\loopa}
\rput[bl](2,3){\loopa}
\rput[bl](3,3){\loopa}
\rput[bl](4,3){\loopa}
\rput[bl](5,3){\loopa}
\rput[bl](6,3){\loopa}
\rput[bl](7,3){\loopa}
\psarc[linecolor=blue,linewidth=2pt](0,1){.5}{90}{270}
\psarc[linecolor=blue,linewidth=2pt](0,3){.5}{90}{270}
\psarc[linecolor=blue,linewidth=2pt](10,1){.5}{-90}{90}
\psarc[linecolor=blue,linewidth=2pt](10,3){.5}{-90}{90}
\psarc[linecolor=purple,linewidth=2pt](1,4){.5}{0}{180}
\psarc[linecolor=purple,linewidth=2pt](7,4){.5}{0}{180}
\psarc[linecolor=purple,linewidth=2pt](4,4){.5}{0}{180}
\psarc[linecolor=purple,linewidth=2pt](7,4){1.5}{0}{180}
\psarc[linecolor=purple,linewidth=2pt](6,3.105){3.62}{14}{166}
\end{pspicture}\nonumber
\eea
\caption{Typical configuration of a logarithmic minimal model. For the ${\cal LM}(1,2)$ model, $\beta=0$ and closed loops are thus forbidden. 
Since the nonlocal degrees of freedom are connectivities, the transfer matrices act on link states which keep track of the planar connectivities. Here the $(r,s)=(1,3)$ boundary condition is applied so that there are $s-1=2$ defects propagating in the bulk.}
\end{figure}
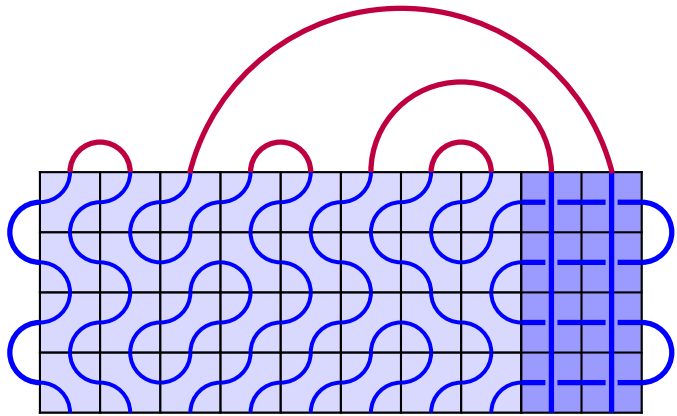

In the continuum scaling limit, the ${\cal LM}(1,p)$ model is described by a logarithmic CFT 
with central charge
\begin{equation}
 c\;=\;  1-6\,\frac{(p-1)^2}{p}\;=\;-2, -7,-\frac{25}{2},-\frac{91}{5},-24,\ldots\qquad p=2,3,4,5,6,\ldots
\label{c}
\end{equation}
and Virasoro conformal weights
\begin{equation}
 \D_{r,s}\ =\ \frac{(rp-s)^{2}-(p-1)^2}{4p},\hspace{1.2cm}r,s\in\mathbb{N}
\label{D}
\end{equation}
The Kac tables of the first two members of this sequence are shown in Figure~2. The ${\cal LM}(1,2)$ model is {\em critical dense polymers}~\cite{PR06} in the Virasoro picture but {\em symplectic fermions}~\cite{Kausch00} in the ${\cal W}$-extended picture.

\subsection{Kac Representations}

Extending ideas originating with Cardy~\cite{Cardy1,Cardy2}, the fusion of representations of the conformal algebra can be implemented on the lattice by combining integrable boundary conditions associated with these representations on the left and right edges of a strip.
On the strip, there is a simple correspondence between integrable boundary conditions, conformal boundary conditions and representations of the chiral algebra. If the chiral algebra is the Virasoro algebra, 
they are each labelled by the Kac labels $(r,s)$ with $r,s\in\mathbb{N}$. 
The integrable boundary conditions are determined by finding solutions to the boundary Yang-Baxter equations (BYBE). As in the rational case~\cite{BP01}, 
these solutions are constructed by fusing integrable seams to the 
boundary as shown in Figure~3. Schematically,
\bea
(r,s)=(r,1)\otimes(1,s)\otimes(1,1),\qquad r,s\in\mathbb{N}
\label{schematic}
\eea
The details of the construction need not concern us here but, for each $r$, there is at least one 
choice for the number of columns 
$\rho-1$ and the integers $k_0$ in the column inhomogeneities $\xi_k=(k+k_0+\half)\lambda$ to yield the required $(r,s)$ integrable boundary.

Although there is a countably infinite number of integrable boundary conditions and corresponding Kac representations $(r,s)$, 
there is no claim that this classification is complete as it is in the rational cases. Indeed, further integrable boundary conditions and corresponding representations can be found by applying additional fusions in (\ref{schematic}). We will exploit this freedom later to construct new boundary conditions associated to
the ${\cal W}$-extended picture.

\psset{unit=.95cm}
\begin{figure}[p]
\begin{center}
\begin{pspicture}(0,0)(7,11)
\rput(3.5,11.5){${\cal LM}(1,2)$}
\psframe[linewidth=0pt,fillstyle=solid,fillcolor=lightlightblue](0,0)(7,11)
\multiput(0,0)(0,2){5}{\psframe[linewidth=0pt,fillstyle=solid,fillcolor=midblue](0,1)(7,2)}
\multirput(1,1)(1,0){6}{\pswedge[fillstyle=solid,fillcolor=red,linecolor=red](0,0){.25}{180}{270}}
\multirput(1,2)(1,0){6}{\pswedge[fillstyle=solid,fillcolor=red,linecolor=red](0,0){.25}{180}{270}}
\multirput(1,2)(0,2){5}{\pswedge[fillstyle=solid,fillcolor=red,linecolor=red](0,0){.25}{180}{270}}
\psframe[linewidth=1.5pt](0,0)(2,2)
\psgrid[gridlabels=0pt,subgriddiv=1]
\rput(.5,10.65){$\vdots$}\rput(1.5,10.65){$\vdots$}\rput(2.5,10.65){$\vdots$}\rput(3.5,10.65){$\vdots$}\rput(4.5,10.65){$\vdots$}\rput(5.5,10.65){$\vdots$}\rput(6.5,10.5){$\vvdots$}
\rput(.5,9.5){$\frac{63}8$}\rput(1.5,9.5){$\frac{35}8$}\rput(2.5,9.5){$\frac{15}8$}\rput(3.5,9.5){$\frac{3}8$}\rput(4.5,9.5){$-\frac 18$}\rput(5.5,9.5){$\frac{3}8$}\rput(6.5,9.5){$\cdots$}
\rput(.5,8.5){$6$}\rput(1.5,8.5){$3$}\rput(2.5,8.5){$1$}\rput(3.5,8.5){$0$}\rput(4.5,8.5){$0$}\rput(5.5,8.5){$1$}\rput(6.5,8.5){$\cdots$}
\rput(.5,7.5){$\frac{35}8$}\rput(1.5,7.5){$\frac {15}8$}\rput(2.5,7.5){$\frac 38$}\rput(3.5,7.5){$-\frac{1}8$}\rput(4.5,7.5){$\frac 38$}\rput(5.5,7.5){$\frac{15}8$}\rput(6.5,7.5){$\cdots$}
\rput(.5,6.5){$3$}\rput(1.5,6.5){$1$}\rput(2.5,6.5){$0$}\rput(3.5,6.5){$0$}\rput(4.5,6.5){$1$}\rput(5.5,6.5){$3$}\rput(6.5,6.5){$\cdots$}
\rput(.5,5.5){$\frac{15}8$}\rput(1.5,5.5){$\frac {3}{8}$}\rput(2.5,5.5){$-\frac 18$}\rput(3.5,5.5){$\frac{3}{8}$}\rput(4.5,5.5){$\frac {15}8$}\rput(5.5,5.5){$\frac{35}{8}$}\rput(6.5,5.5){$\cdots$}
\rput(.5,4.5){$1$}\rput(1.5,4.5){$0$}\rput(2.5,4.5){$0$}\rput(3.5,4.5){$1$}\rput(4.5,4.5){$3$}\rput(5.5,4.5){$6$}\rput(6.5,4.5){$\cdots$}
\rput(.5,3.5){$\frac 38$}\rput(1.5,3.5){$-\frac 18$}\rput(2.5,3.5){$\frac 38$}\rput(3.5,3.5){$\frac{15}8$}\rput(4.5,3.5){$\frac{35}8$}\rput(5.5,3.5){$\frac{63}8$}\rput(6.5,3.5){$\cdots$}
\rput(.5,2.5){$0$}\rput(1.5,2.5){$0$}\rput(2.5,2.5){$1$}\rput(3.5,2.5){$3$}\rput(4.5,2.5){$6$}\rput(5.5,2.5){$10$}\rput(6.5,2.5){$\cdots$}
\rput(.5,1.5){$-\frac 18$}\rput(1.5,1.5){$\frac 38$}\rput(2.5,1.5){$\frac{15}8$}\rput(3.5,1.5){$\frac{35}8$}\rput(4.5,1.5){$\frac{63}8$}\rput(5.5,1.5){$\frac{99}8$}\rput(6.5,1.5){$\cdots$}
\rput(.5,.5){$0$}\rput(1.5,.5){$1$}\rput(2.5,.5){$3$}\rput(3.5,.5){$6$}\rput(4.5,.5){$10$}\rput(5.5,.5){$15$}\rput(6.5,.5){$\cdots$}
{\color{blue}
\rput(.5,-.5){$1$}
\rput(1.5,-.5){$2$}
\rput(2.5,-.5){$3$}
\rput(3.5,-.5){$4$}
\rput(4.5,-.5){$5$}
\rput(5.5,-.5){$6$}
\rput(6.5,-.5){$r$}
\rput(-.5,.5){$1$}
\rput(-.5,1.5){$2$}
\rput(-.5,2.5){$3$}
\rput(-.5,3.5){$4$}
\rput(-.5,4.5){$5$}
\rput(-.5,5.5){$6$}
\rput(-.5,6.5){$7$}
\rput(-.5,7.5){$8$}
\rput(-.5,8.5){$9$}
\rput(-.5,9.5){$10$}
\rput(-.5,10.5){$s$}}
\end{pspicture}\qquad\qquad
\begin{pspicture}(0,0)(7,11)
\rput(3.5,11.5){${\cal LM}(1,3)$}
\psframe[linewidth=0pt,fillstyle=solid,fillcolor=lightlightblue](0,0)(7,11)
\multiput(0,1)(0,3){3}{\psframe[linewidth=0pt,fillstyle=solid,fillcolor=midblue](0,1)(7,2)}
\multirput(1,1)(1,0){6}{\pswedge[fillstyle=solid,fillcolor=red,linecolor=red](0,0){.25}{180}{270}}
\multirput(1,2)(1,0){6}{\pswedge[fillstyle=solid,fillcolor=red,linecolor=red](0,0){.25}{180}{270}}
\multirput(1,3)(1,0){6}{\pswedge[fillstyle=solid,fillcolor=red,linecolor=red](0,0){.25}{180}{270}}
\multirput(1,6)(0,3){2}{\pswedge[fillstyle=solid,fillcolor=red,linecolor=red](0,0){.25}{180}{270}}
\psframe[linewidth=1.5pt](0,0)(2,3)
\psgrid[gridlabels=0pt,subgriddiv=1]
\rput(.5,10.65){$\vdots$}\rput(1.5,10.65){$\vdots$}\rput(2.5,10.65){$\vdots$}\rput(3.5,10.65){$\vdots$}\rput(4.5,10.65){$\vdots$}\rput(5.5,10.65){$\vdots$}\rput(6.5,10.5){$\vvdots$}
\rput(.5,9.5){$\frac{15}4$}\rput(1.5,9.5){$1$}\rput(2.5,9.5){$-\frac 14$}\rput(3.5,9.5){$0$}\rput(4.5,9.5){$\frac 74$}\rput(5.5,9.5){$5$}\rput(6.5,9.5){$\cdots$}
\rput(.5,8.5){$\frac 83$}\rput(1.5,8.5){$\frac 5{12}$}\rput(2.5,8.5){$-\frac 13$}\rput(3.5,8.5){$\frac 5{12}$}\rput(4.5,8.5){$\frac 83$}\rput(5.5,8.5){$\frac{77}{12}$}\rput(6.5,8.5){$\cdots$}
\rput(.5,7.5){$\frac 74$}\rput(1.5,7.5){$0$}\rput(2.5,7.5){$-\frac 14$}\rput(3.5,7.5){$1$}\rput(4.5,7.5){$\frac {15}4$}\rput(5.5,7.5){$8$}\rput(6.5,7.5){$\cdots$}
\rput(.5,6.5){$1$}\rput(1.5,6.5){$-\frac 14$}\rput(2.5,6.5){$0$}\rput(3.5,6.5){$\frac 74$}\rput(4.5,6.5){$5$}\rput(5.5,6.5){$\frac{39}4$}\rput(6.5,6.5){$\cdots$}
\rput(.5,5.5){$\frac 5{12}$}\rput(1.5,5.5){$-\frac 13$}\rput(2.5,5.5){$\frac 5{12}$}\rput(3.5,5.5){$\frac 83$}\rput(4.5,5.5){$\frac {77}{12}$}\rput(5.5,5.5){$\frac{35}{3}$}\rput(6.5,5.5){$\cdots$}
\rput(.5,4.5){$0$}\rput(1.5,4.5){$-\frac 14$}\rput(2.5,4.5){$1$}\rput(3.5,4.5){$\frac {15}4$}\rput(4.5,4.5){$8$}\rput(5.5,4.5){$\frac{55}4$}\rput(6.5,4.5){$\cdots$}
\rput(.5,3.5){$-\frac 14$}\rput(1.5,3.5){$0$}\rput(2.5,3.5){$\frac 74$}\rput(3.5,3.5){$5$}\rput(4.5,3.5){$\frac{39}4$}\rput(5.5,3.5){$16$}\rput(6.5,3.5){$\cdots$}
\rput(.5,2.5){$-\frac 13$}\rput(1.5,2.5){$\frac 5{12}$}\rput(2.5,2.5){$\frac 83$}\rput(3.5,2.5){$\frac{77}{12}$}\rput(4.5,2.5){$\frac{35}{3}$}\rput(5.5,2.5){$\frac{221}{12}$}\rput(6.5,2.5){$\cdots$}
\rput(.5,1.5){$-\frac 14$}\rput(1.5,1.5){$1$}\rput(2.5,1.5){$\frac{15}4$}\rput(3.5,1.5){$8$}\rput(4.5,1.5){$\frac{55}4$}\rput(5.5,1.5){$21$}\rput(6.5,1.5){$\cdots$}
\rput(.5,.5){$0$}\rput(1.5,.5){$\frac 74$}\rput(2.5,.5){$5$}\rput(3.5,.5){$\frac{39}4$}\rput(4.5,.5){$16$}\rput(5.5,.5){$\frac{95}4$}\rput(6.5,.5){$\cdots$}
{\color{blue}
\rput(.5,-.5){$1$}
\rput(1.5,-.5){$2$}
\rput(2.5,-.5){$3$}
\rput(3.5,-.5){$4$}
\rput(4.5,-.5){$5$}
\rput(5.5,-.5){$6$}
\rput(6.5,-.5){$r$}
\rput(-.5,.5){$1$}
\rput(-.5,1.5){$2$}
\rput(-.5,2.5){$3$}
\rput(-.5,3.5){$4$}
\rput(-.5,4.5){$5$}
\rput(-.5,5.5){$6$}
\rput(-.5,6.5){$7$}
\rput(-.5,7.5){$8$}
\rput(-.5,8.5){$9$}
\rput(-.5,9.5){$10$}
\rput(-.5,10.5){$s$}}
\end{pspicture}
\end{center}
\caption{Extended Kac table of conformal weights $\Delta_{r,s}$ for ${\cal LM}(1,2)$ and ${\cal LM}(1,3)$ with $c=-2, -7$ respectively. 
In general, the entries relate to distinct Kac representations $(r,s)$ even if the conformal weights coincide.  
The periodicity $\Delta_{r,s}=\Delta_{r+1,s+p}$ is made manifest by the shading of the rows.
The Kac representations which also happen to be irreducible representations are marked with a red shaded quadrant in the top-right corner. 
Pairs of irreducible representations are identified according to $(r,2)\equiv (1,2r)$ and $(r,3)\equiv (1,3r)$. 
The heavy frames in the lower-left corners delimit the Kac tables in the ${\cal W}$-extended picture.
}
\end{figure}

\begin{figure}[htbp]
\setlength{\unitlength}{12mm}
\psset{unit=12mm}
\bea
\mbox{}\qquad
\raisebox{-1.4\unitlength}[1.6\unitlength][1.2\unitlength]
{\begin{picture}(5.,2.0)
\put(1,.5){\color{lightlightblue}\rule{5\unitlength}{2\unitlength}}
\put(0.45,1.5){\makebox(0,0)[]{$=$}}
\put(.45,2.83){\spos{bc}{\color{blue}=}}
\put(-.1,2.8){\spos{bc}{\color{blue}(r,s)}}
\put(3.9,2.8){\spos{bc}{\color{blue}(r,1)}}
\put(6,2.8){\spos{bc}{\color{blue}\otimes}}
\put(-0.1,0.5){\line(0,1){2}}
\multiput(1,0.5)(1,0){4}{\line(0,1){2}}
\multiput(5,0.5)(1,0){2}{\line(0,1){2}}
\multiput(1,0.5)(0,1){3}{\line(1,0){5}}
\put(-0.6,1.5){\line(1,2){0.5}}
\put(-0.6,1.5){\line(1,-2){0.5}}
\put(5,.5){\oval(.2,.2)[tr]}
\put(5,1.5){\oval(.2,.2)[tr]}
\multiput(1,.5)(1,0){2}{\oval(.2,.2)[tr]}
\multiput(1,1.5)(1,0){2}{\oval(.2,.2)[tr]}
\put(1.5,1){\spos{}{u\!-\!\xi_{\rho\!-\!1}}}
\put(2.5,1){\spos{}{u\!-\!\xi_{\rho\!-\!2}}}
\put(5.5,1){\spos{}{u\!-\!\xi_1}}
\put(1.5,2){\spos{}{-\!u\!-\!\xi_{\rho\!-\!2}}}
\put(2.5,2){\spos{}{-\!u\!-\!\xi_{\rho\!-\!3}}}
\put(5.5,2){\spos{}{-\!u\!-\!\xi_0}}
\put(-0.31,1.5){\spos{}{u}}
\multiput(0,0)(1,0){5}{\psarc[linewidth=2pt](.5,2.5){1}{0}{40}}
\multiput(0,0)(1,0){5}{\psline[linewidth=2pt](1.5,.3)(1.5,.5)}
\psline[linewidth=2pt](.8,1)(1,1)
\psline[linewidth=2pt](.8,2)(1,2)
\psline[linewidth=2pt](-.55,1)(-.35,1)
\psline[linewidth=2pt](-.55,2)(-.35,2)
\end{picture}}
\raisebox{-1.4\unitlength}[1.6\unitlength][1.2\unitlength]
{\begin{pspicture}(4.5,2.0)
\put(1,.5){\color{lightpurple}\rule{5\unitlength}{2\unitlength}}
\psline[linewidth=2pt](1,.5)(1,2.5)
\put(2.9,2.8){\spos{bc}{\color{blue}(1,s)}}
\put(6.5,2.8){\spos{bc}{\color{blue}(1,1)}}
\put(6,2.8){\spos{bc}{\color{blue}\otimes}}
\put(6.5,0.5){\line(0,1){2}}
\multiput(1,0.5)(1,0){4}{\line(0,1){2}}
\multiput(5,0.5)(1,0){2}{\line(0,1){2}}
\multiput(1,0.5)(0,1){3}{\line(1,0){5}}
\put(6,1.5){\line(1,2){0.5}}
\put(6,1.5){\line(1,-2){0.5}}
\put(5,.5){\oval(.2,.2)[tr]}
\put(5,1.5){\oval(.2,.2)[tr]}
\multiput(1,.5)(1,0){2}{\oval(.2,.2)[tr]}
\multiput(1,1.5)(1,0){2}{\oval(.2,.2)[tr]}
\multiput(0,0)(1,0){5}{\psarc[linewidth=2pt](.5,2.5){1}{0}{40}}
\multiput(0,0)(1,0){5}{\psline[linewidth=2pt](1.5,.3)(1.5,2.5)}
\multiput(0,0)(0,1){2}{\psline[linewidth=2pt](1,1)(1.4,1)}
\multiput(0,0)(0,1){2}{\psline[linewidth=2pt](5.6,1)(6,1)}
\multiput(0,0)(0,1){2}{\multiput(0,0)(1,0){4}{\psline[linewidth=2pt](1.6,1)(2.4,1)}}
\multiput(6,0.5)(0,2){2}{\makebox(0.5,0){\dotfill}}
\psarc[linewidth=2pt](5.7,1.5){.6}{-61}{61}
\end{pspicture}}
\qquad\qquad\qquad\nonumber\\[-6pt]
\mbox{}\hspace{-2\unitlength}
\underbrace{\hspace{4.95\unitlength}}_{\mbox{\small $\rho-1$ columns}}\underbrace{\hspace{4.95\unitlength}}_{\mbox{\small $s-1$ columns}}\;\quad\nonumber
\eea
\caption{Construction of the integrable boundary condition corresponding to the Kac representation $(r,s)$. 
The $(r,s)$ solution to the BYBE is built by fusing integrable $(r,1)$ and $(1,s)$ seams to the 
$(1,1)$ or vacuum boundary. The column inhomogeneities are $\xi_k=(k+k_0+\half)\lambda$.
There is at least one choice of the integers $\rho$ and $k_0$ for each $r$.
The $\rho+s-2$ columns are considered part of the right boundary. 
The arches at the top close to the left with up to $\rho+s-2$ defects propagating in the bulk. 
Some of the $s$-arches can close with some of the $r$-arches. 
Left boundary solutions $(r',s')$ are constructed similarly.}
\end{figure}
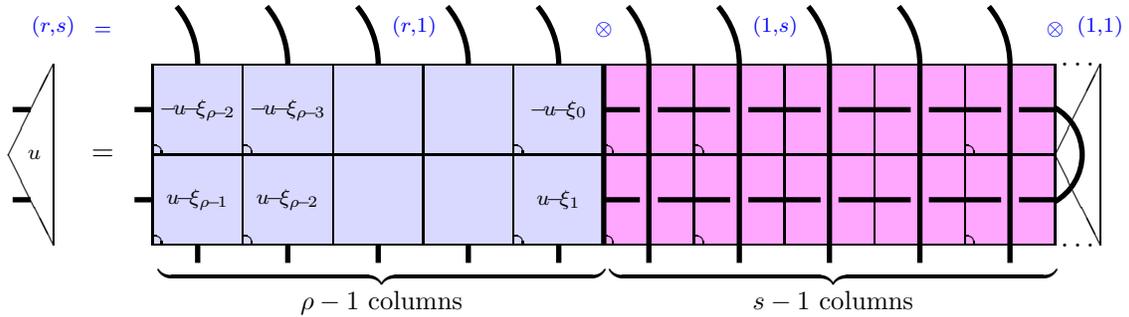

For the ${\cal LM}(1,p)$ models, the conformal character of the Kac representation $(r,s)$ is given by
\begin{equation}
 \chit_{r,s}(q)\ =\ \frac{q^{\frac{1-c}{24}+\D_{r,s}}}{\eta(q)}\big(1-q^{rs}\big)
  \ =\ \frac{1}{\eta(q)}\big(q^{(rp-s)^2/4p}-q^{(rp+s)^2/4p}\big)
\label{chikac}
\end{equation}
where the Dedekind eta function is defined by
\begin{equation}
 \eta(q)\ =\ q^{1/24}\prod_{m=1}^\infty(1-q^m)
\label{eta}
\end{equation}
A priori, a Kac representation is either irreducible or reducible.
The only Kac representations to be used here are the {\em irreducible} Kac representations 
\begin{equation}
 \{(1,kp),(r,s);\ s\in\mathbb{Z}_{1,p};\ r,k\in\mathbb{N}\}
\label{Kacirr}
\end{equation}
Since their characters
all correspond to irreducible Virasoro characters, these Kac representations must indeed
themselves be irreducible. The set (\ref{Kacirr}) constitutes an 
exhaustive list of irreducible Kac representations.
Two Kac representations are naturally identified if they have identical
conformal weights and are both irreducible. For the ${\cal LM}(1,p)$ models, the relations
\begin{equation}
 (1,kp)\ \equiv\ (k,p),\qquad k\in\mathbb{N}
\label{idirr}
\end{equation}
are the only such identifications.

\subsection{Indecomposable Representations of Rank 2}

Indecomposable representations of rank 2 or 3 arise from certain fusions
of Kac representations in ${\cal LM}(p,p')$ and are generally denoted $\R_{r,s}^{a,a'}$.
In the case of ${\cal LM}(1,p)$, there are no rank-3 representations and
the rank-2 representations are here given by the shorthand notation
\begin{equation}
 \R_r^a\ \equiv\  \R_{1,rp}^{0,a}\ \equiv\ \R_{r,p}^{0,a},\qquad r\in\mathbb{N};\ a\in\mathbb{Z}_{1,p-1}
\end{equation}
Their characters read
\begin{equation}
 \chit[\R_r^a](q)\ =\ \chit_{r,p-a}(q)+\chit_{r,p+a}(q)\ =\ \chit_{1,rp-a}(q)+\chit_{1,rp+a}(q)
   \ =\ \chit_{r-1,a}(q)+2\chit_{r,p-a}(q)+\chit_{r+1,a}(q)
\end{equation}

\subsection{\protect\boldmath Virasoro Fusion Algebra of ${\cal LM}(1,p)$}

The fundamental fusion algebra of ${\cal LM}(1,p)$ \cite{RP0707}
\begin{equation}
 \big\langle(2,1),(1,2)\big\rangle_{1,p}\ =\  \big\langle(r,s),\R_{r}^{a};\ 
  a\in\mathbb{Z}_{1,p-1}; \ s\in\mathbb{Z}_{1,p};\ r\in\mathbb{N}\big\rangle_{1,p} 
\label{fund1p}
\end{equation}
is commutative and associative. To describe the fusion rules, we
set $\R_{n}^0\equiv(n,p)$, $\R_{-1}^a\equiv-\R_1^a$, $\R_0^a\equiv0$ and 
\bea
 \dnn\!\!&=&\!\!  2-\delta_{j,|n-n'|}-\delta_{j,n+n'}  \nn
 \ddnn\!\!&=&\!\!   4-3\delta_{j,|n-n'|-1}-2\delta_{j,|n-n'|}-\delta_{j,|n-n'|+1}
   -\delta_{j,n+n'-1}-2\delta_{j,n+n'}-3\delta_{j,n+n'+1}
\label{d24}
\eea
For $a,a',s_0,s'_0\in\mathbb{Z}_{1,p-1}$, the fusion rules \cite{RP0707} are
\bea
 (1,s_0)\otimes\R_{n}^0\!\!&=&\!\! \bigoplus_{i=0}^{\lfloor\frac{s_0-1}{2}\rfloor}\ \R_{n}^{s_0-1-2i}    \nn
 \R_{n}^{0}\otimes\R_{n'}^{0}\!\!&=&\!\! \bigoplus_{j=|n-n'|+1,\ \!{\rm by}\ \!2}^{n+n'-1}
   \Big\{\bigoplus_{i=0}^{\lfloor\frac{p-1}{2}\rfloor}\R_{j}^{p-1-2i}\Big\}\nn
 \R_{n}^0\otimes\R_{n'}^{a'}\!\!&=&\!\!
   \Big(\bigoplus_{j=|n-n'|,\ \!{\rm by}\ \!2}^{n+n'}\dnn\Big\{\bigoplus_{i=0}^{\lfloor\frac{a'-1}{2}\rfloor}
     \R_{j}^{a'-1-2i}\Big\}\Big)
  \oplus\Big(\bigoplus_{j=|n-n'|+1,\ \!{\rm by}\ \!2}^{n+n'-1}2\ \! \Big\{
    \bigoplus_{i=0}^{\lfloor\frac{p-a'-1}{2}\rfloor}\R_{j}^{p-a'-1-2i}\Big\}\Big)\nn
    \label{VirFusion1}
\eea
where for $s_0+s'_0,s_0+a,a+a'\leq p$
\bea
 (1,s_0)\otimes(1,s'_0)\!\!&=&\!\! \bigoplus_{j=|s_0-s'_0|+1,\ \!{\rm by}\ \!2}^{s_0+s'_0-1}(1,j)   \nn
 (1,s_0)\otimes\R_{n}^a\!\!&=&\!\!  
  \Big\{\bigoplus_{i=0}^{\min\{s_0-1,\lfloor\frac{s_0+a-1}{2}\rfloor\}}\R_{n}^{s_0+a-1-2i}\Big\}
   \oplus\Big\{\bigoplus_{i=0}^{\lfloor\frac{s_0-a-1}{2}\rfloor}\R_{n}^{s_0-a-1-2i}\Big\}   \nn
 \R_{n}^a\otimes\R_{n'}^{a'} \!\!&=&\!\!
   \Big(
     \bigoplus_{j=|n-n'|,\ \!{\rm by}\ \!2}^{n+n'}\dnn
      \Big\{\big(\bigoplus_{i=0}^{\lfloor\frac{|a-a'|-1}{2}\rfloor}\R_{j}^{|a-a'|-1-2i}\Big)     
         \oplus\Big(\bigoplus_{i=0}^{\lfloor\frac{a+a'-1}{2}\rfloor}\R_{j}^{a+a'-1-2i}\Big)\Big\}\Big)\nn
 \!\!&\oplus&\!\!
   \Big(\bigoplus_{j=|n-n'|+1,\ \!{\rm by}\ \!2}^{n+n'-1}
    2\ \! \Big\{\Big(\bigoplus_{i=0}^{\lfloor\frac{p-|a-a'|-1}{2}\rfloor}\R_{j}^{p-|a-a'|-1-2i}\Big)
    \oplus\Big(\bigoplus_{i=0}^{\lfloor\frac{p-a-a'-1}{2}\rfloor}\R_{j}^{p-a-a'-1-2i}\Big)\Big\}\Big)\nn
       \label{VirFusion2}
\eea
while for $s_0+s'_0,s_0+a,a+a'>p$
\bea
 (1,s_0)\otimes(1,s'_0)\!\!&=&\!\! 
   \Big(\bigoplus_{j=|s_0-s'_0|+1,\ \!{\rm by}\ \!2}^{2p-s_0-s'_0-1}(1,j)\Big)
  \oplus\Big\{ \bigoplus_{i=0}^{\lfloor\frac{s_0+s'_0-p-1}{2}\rfloor} \R_{1}^{s_0+s'_0-p-1-2i}\Big\}   \nn
 (1,s_0)\otimes\R_{n}^a\!\!&=&\!\! \Big\{
   \bigoplus_{i=0}^{\lfloor\frac{s_0+a-p-1}{2}\rfloor}\Big(\R_{n-1}^{s_0+a-p-1-2i}
     \oplus\R_{n+1}^{s_0+a-p-1-2i}\Big)     \Big\}\nn
 \!\!&\oplus&\!\!
   \Big\{\bigoplus_{i=0}^{\min\{p-a-1,\lfloor\frac{2p-s_0-a-1}{2}\rfloor\}}\R_{n}^{2p-s_0-a-1-2i}\Big\}
    \oplus\Big\{\bigoplus_{i=0}^{\lfloor\frac{s_0-a-1}{2}\rfloor}\R_{n}^{s_0-a-1-2i}\Big\}\nn
 \R_{n}^a\otimes\R_{n'}^{a'}\!\!&=&\!\!
  \Big(\bigoplus_{j=|n-n'|-1,\ \!{\rm by}\ \!2}^{n+n'+1}\ddnn
    \Big\{\bigoplus_{i=0}^{\lfloor\frac{a+a'-p-1}{2}\rfloor}\R_{j}^{a+a'-p-1-2i}\Big\}\Big)\nn
 \!\!&\oplus&\!\!
   \Big(
     \bigoplus_{j=|n-n'|,\ \!{\rm by}\ \!2}^{n+n'}\dnn
      \Big\{\Big(\bigoplus_{i=0}^{\lfloor\frac{|a-a'|-1}{2}\rfloor}\R_{j}^{|a-a'|-1-2i}\Big)     
         \oplus\Big(\bigoplus_{i=0}^{\lfloor\frac{2p-a-a'-1}{2}\rfloor}\R_{j}^{2p-a-a'-1-2i}\Big)\Big\}\Big)\nn
 \!\!&\oplus&\!\!
   \Big(\bigoplus_{j=|n-n'|+1,\ \!{\rm by}\ \!2}^{n+n'-1}
    2\ \! \Big\{\bigoplus_{i=0}^{p-\max\{a,a'\}-1}\R_{j}^{p-|a-a'|-1-2i}\Big\}\Big)
   \label{VirFusion3}
\eea

In the special case of critical dense polymers ${\cal LM}(1,2)$, the fundamental fusion algebra reads
\begin{equation}
  \big\langle(2,1),(1,2)\big\rangle_{1,2}\ =\ \big\langle(r,s),\R_{r}^{1};\ 
    r\in\mathbb{N},\ s\in\mathbb{Z}_{1,2}\big\rangle_{1,2} 
\end{equation}
with fusion rules
\bea
\begin{array}{rcl}
\disp (r,1)\otimes(r',s)\!\!&=&\!\!   \disp\bigoplus_{j=|r-r'|+1,\ \!{\rm by}\ \!2}^{r+r'-1} (j,s)\\
\disp  (r,2)\otimes(r',2)\!\!&=&\!\!   \disp\bigoplus_{j=|r-r'|+1,\ \!{\rm by}\ \!2}^{r+r'-1}\R_{j}^{1}   \\
\disp (r,1)\otimes\R_{r'}^{1}\!\!&=&\!\!  \disp\bigoplus_{j=|r-r'|+1,\ \!{\rm by}\ \!2}^{r+r'-1}\R_{j}^{1}   \\
\disp (r,2)\otimes\R_{r'}^{1}\!\!&=&\!\! \disp\bigoplus_{j=|r-r'|}^{r+r'}\drr(j,2)   \\
\disp \R_{r}^{1}\otimes\R_{r'}^{1}\!\!&=&\!\! \disp\bigoplus_{j=|r-r'|}^{r+r'}\drr\R_{j}^{1}  
 \end{array}
\label{VirFusionRules}
\eea
Here the superscript $1$ on $\R_{k}^1$ is redundant but kept for consistency of notation.

Introducing 
\begin{equation}
 \epsilon(n)=\frac{1}{2}\big(1-(-1)^n\big),\qquad n\in\mathbb{Z}
\label{epsn}
\end{equation}
a particular subset of the fusion rules 
(\ref{VirFusion1}), (\ref{VirFusion2}) and (\ref{VirFusion3}) can be written in the following compact form
\bea
 (1,s)\otimes(1,s')&=&\Big(\bigoplus_{j=|s-s'|+1,\ \!{\rm by}\ \!2}^{p-|p-s-s'|-1}(1,j)\Big)
   \oplus\Big(\bigoplus_{\ell=\epsilon(s+s'-p-1),\ \!{\rm by}\ \!2}^{s+s'-p-1}\R_1^\ell\Big)\nn
 (1,s)\otimes\R_1^a&=&\Big(\bigoplus_{\ell=|s-a|+1,\ \!{\rm by}\ \!2}^{p-|p-s-a|-1}\R_1^\ell\Big)
   \oplus2\Big(\bigoplus_{\ell=\epsilon(s-a-1),\ \!{\rm by}\ \!2}^{s-a-1}\R_1^\ell\Big)
   \oplus\Big(\bigoplus_{\ell=\epsilon(s+a-p-1),\ \!{\rm by}\ \!2}^{s+a-p-1}\R_2^\ell\Big)\nn
 \R_1^a\otimes\R_1^{a'}&=&2\Big(\bigoplus_{\ell=|p-a-a'|+1,\ \!{\rm by}\ \!2}^{p-|a-a'|-1}\R_1^\ell\Big)
   \oplus4\Big(\bigoplus_{\ell=\epsilon(p-a-a'-1),\ \!{\rm by}\ \!2}^{p-a-a'-1}\R_1^\ell\Big)
   \oplus\Big(\bigoplus_{\ell=\epsilon(a+a'-p-1),\ \!{\rm by}\ \!2}^{a+a'-p-1}
    \big(\R_1^\ell\oplus\R_3^\ell\big)\Big)\nn
  &\oplus&\Big(\bigoplus_{\ell=|a-a'|+1,\ \!{\rm by}\ \!2}^{p-|p-a-a'|-1}\R_2^\ell\Big)
    \oplus2\Big(\bigoplus_{\ell=\epsilon(a+a'+1),\ \!{\rm by}\ \!2}^{|a-a'|-1}\R_2^\ell\Big)
\label{VirFusCompact}
\eea
where $s,s'\in\mathbb{Z}_{1,p}$ and $a,a'\in\mathbb{Z}_{1,p-1}$.
These expressions correspond to setting $n,n'=1$ in (\ref{VirFusion1}),
(\ref{VirFusion2}) and (\ref{VirFusion3}) and will be used below.

\section{\protect\boldmath Logarithmic Minimal Model ${\cal LM}(1,p)$: Extended Picture}

We now consider the logarithmic minimal models ${\cal LM}(1,p)$ as `rational' logarithmic CFTs 
with extended $W$ symmetry. In this picture, the characters of the infinity of Kac representations 
are reorganized into a {\em finite} number of extended characters given as infinite sums of the Kac 
characters with suitable multiplicities. 
The central charges are given by (\ref{c}) but now there is a {\em finite} Kac table of conformal weights 
as shown in Figure~2
\begin{equation}
 \D_{r,s}\ =\ \frac{(rp-s)^{2}-(p-1)^2}{4p},\qquad r\in\mathbb{Z}_{1,2};\ \ s\in\mathbb{Z}_{1,p}
\label{extConfWeights}
\end{equation}
corresponding to the $2p$ irreducible representations here denoted by
\bea
\ketw{r,s},\qquad r\in\mathbb{Z}_{1,2};\quad s\in\mathbb{Z}_{1,p}
\eea
Explicitly, the characters of these irreducible representations are given by
\bea
 \chih_{1,s}(q)&=&\chit[\ketw{1,s}]
    \ =\ \frac{1}{\eta(q)}\Big(\frac{s}{p}\,\vartheta_{p-s,p}(q)+2\vartheta'_{p-s,p}(q)\Big)\nn
 \chih_{2,s}(q)&=&\chit[\ketw{2,s}] 
   \ =\ \frac{1}{\eta(q)}\Big(\frac{s}{p}\,\vartheta_{s,p}(q)-2\vartheta'_{s,p}(q)\Big)
\eea 
where $\eta(q)$ is the Dedekind eta function (\ref{eta}) and $\vartheta_{s,p}(q)=\vartheta_{s,p}(q,1)$ with
\begin{equation}
 \vartheta_{s,p}(q,z)=\sum_{n\in {\mathbb Z}+\frac{s}{2p}} q^{n^2p} z^n,
   \qquad \vartheta'(q)=z\,\frac{\partial}{\partial z}\vartheta_{s,p}(q,z)\Big|_{z=1}
\end{equation}

The $2p$ irreducible characters in the extended picture of  
${\cal LM}(1,p)$ can be expanded in terms of the characters of the Kac representations as follows
\bea
  \chih_{1,s}(q)&=&\sum_{n=1}^\infty (2n-1)\,\chit_{2n-1,s}(q)\nn
  \chih_{2,s}(q)&=&\sum_{n=1}^\infty 2n\,\chit_{2n,s}(q)
\label{extchars}
\eea
where $s\in\mathbb{Z}_{1,p}$.
In addition, there are $2p-2$ indecomposable rank-2 representations which we denote by
\begin{equation}
 \ketw{\R_1^a},\quad \ketw{\R_{2}^a},\qquad a\in\mathbb{Z}_{1,p-1}
\end{equation}
Though inequivalent, their characters are equal in pairs
\begin{equation}
 \chit[\ketw{\R_1^a}](q) = \chit[\ketw{\R_{2}^{p-a}}](q) = 2 \chih_{1,p-a}(q) + 2 \chih_{2,a}(q)
\end{equation}
As we will discuss below, the rank-2 representations themselves differ 
in their Jordan-cell structures and general embedding structures.

The complete set of $4p-2$ representations is believed to close under fusion, with conjectured 
fusion rules given in \cite{GK9606,FK07,GR07}. 
In our notation and with $\ketw{\R_r^0}\equiv\ketw{r,p}$, this conjectured fusion algebra reads 
\bea
 \ketw{r,s}\,\hat\otimes\,\ketw{r',s'}&=&\Big(\bigoplus_{j=|s-s'|+1,\ \!{\rm by}\ \!2}^{p-|p-s-s'|-1}
   \ketw{r\cdot r',j}\Big)
    \oplus\Big(\bigoplus_{\ell=\epsilon(s+s'-p-1),\ \!{\rm by}\ \!2}^{s+s'-p-1}
    \ketw{\R_{r\cdot r'}^\ell}\Big)\nn
 \ketw{r,s}\,\hat\otimes\,\ketw{\R_{r'}^a}&=&\Big(\bigoplus_{\ell=|s-a|+1,\ \!{\rm by}\ \!2}^{p-|p-s-a|-1}
   \ketw{\R_{r\cdot r'}^\ell}\Big)
   \oplus2\Big(\bigoplus_{\ell=\epsilon(s-a-1),\ \!{\rm by}\ \!2}^{s-a-1}\ketw{\R_{r\cdot r'}^\ell}\Big)\nn
  &\oplus&2\Big(\bigoplus_{\ell=\epsilon(s+a-p-1),\ \!{\rm by}\ \!2}^{s+a-p-1}
    \ketw{\R_{3-r\cdot r'}^\ell}\Big)\nn
 \ketw{\R_{r}^a}\,\hat\otimes\,\ketw{\R_{r'}^{a'}}
  &=&2\Big(\ketw{r,p-a}\oplus\ketw{3-r,a}\Big)\,\hat\otimes\,\ketw{\R_{r'}^{a'}}
\label{extFusion} 
\eea
where $r,r'\in\mathbb{Z}_{1,2}$ and
\begin{equation}
 r\cdot r'=1+\epsilon(r+r')
\label{rdotr}
\end{equation}

\subsection{Extended Boundary Conditions and their Fusion Algebra}

The extended vacuum character of ${\cal LM}(1,p)$ is
\begin{equation}
 \chih_{1,1}(q)=\sum_{n=1}^\infty (2n-1)\,\chit_{2n-1,1}(q)
\label{chih11}
\end{equation}
This suggests that the corresponding integrable boundary condition should be given by the direct sum
\begin{equation}
 \ketw{1,1}=\bigoplus_{n=1}^\infty \,(2n-1)\,(2n-1,1)
\end{equation}
However, the BYBE is {\em not} linear and sums of solutions do {\em not} usually give new solutions. 
Rather, the BYBE is closed under fusions. It follows that if we can construct the desired direct sum from fusions, then automatically it will be a solution of the BYBE.

We thus consider the triple fusion
\bea
 (2n-1,1)\otimes(2n-1,1)\otimes(2n-1,1)&=&\bigoplus_{k=1}^{n}(2k-1)(2k-1,1)
   \oplus\bigoplus_{k=n+1}^{3n-2}(3n-k-1)(2k-1,1)\nn
 &=&(1,1)\oplus3(3,1)\oplus5(5,1)\oplus\cdots\oplus (2n-1)(2n-1,1)\oplus\cdots   \nn
\eea
The coefficients in the tail for $(2n+1,1)$ and beyond have not saturated for this finite $n$, but as $n\to\infty$ the coefficients progressively stabilize and exactly reproduce the multiplicities of the extended vacuum $\ketw{1,1}$. We conclude that the extended vacuum boundary condition can be 
constructed by fusing three $r$-type integrable seams to the boundary
\begin{equation}
 \ketw{1,1}:=\lim_{n\to\infty}(2n-1,1)\otimes(2n-1,1)\otimes(2n-1,1)
  =\bigoplus_{n=1}^\infty \,(2n-1)\,(2n-1,1)
\end{equation}
This extended vacuum has the following remarkable stability property
\begin{equation}
  (2m-1,1)\otimes \ketw{1,1}=(2m-1)\,\Big(\bigoplus_{n=1}^\infty \,(2n-1)\,(2n-1,1)\Big)
   =(2m-1)\, \ketw{1,1}
\label{prop11}
\end{equation}
and more generally satisfies
\bea
 (2m-1,s)\otimes \ketw{1,1}&=&(2m-1)\,\Big(\bigoplus_{n=1}^\infty \,(2n-1)\,(2n-1,s)\Big)\nn
 (2m,s)\otimes \ketw{1,1}&=&2m\,\Big(\bigoplus_{n=1}^\infty \,2n\,(2n,s)\Big)\nn
 \calR_{2m-1}^a\otimes\ketw{1,1}&=&(2m-1)\,\Big(\bigoplus_{n=1}^\infty \,(2n-1)\,\calR_{2n-1}^a\Big)\nn
 \calR_{2m}^a\otimes\ketw{1,1}&=&2m\,\Big(\bigoplus_{n=1}^\infty \,2n\,\calR_{2n}^a\Big)
\label{prop11gen}
\eea
for all $s\in\mathbb{Z}_{1,p}$, $a\in\mathbb{Z}_{1,p-1}$ and $m\in\mathbb{N}$.

We want the extended vacuum boundary representation $\ketw{1,1}$ to act as the 
identity in the fusion algebra associated to the extended picture. In particular, we 
require that
\begin{equation}
 \ketw{1,1}\, \hat\otimes\, \ketw{1,1}=\ketw{1,1}
\end{equation}
where $\hat\otimes$ denotes the fusion multiplication in the extended picture.
This can be achieved by interpreting this extended fusion multiplication as
a limit of a rescaled fusion in the logarithmic minimal model  
\begin{equation}
 \ketw{1,1}\, \hat\otimes\, \ketw{1,1}:=\lim_{n\to\infty}\Big(\frac{1}{(2n-1)^3}(2n-1,1)\otimes(2n-1,1)
   \otimes(2n-1,1)\otimes\ketw{1,1}\Big)=\ketw{1,1}
\end{equation}
thus ensuring that fusion in the extended picture has a natural implementation on the lattice.

Now, a representation in the extended picture $\ketw{\hat A}$ is constructed as the integrable
boundary condition $A\otimes\ketw{1,1}$ where $A$ is some representation in the logarithmic
minimal model. Fusion in the extended picture is then described by
\bea
 \ketw{\hat A}\, \hat\otimes\, \ketw{\hat B}&=&
  \Big(A\otimes \ketw{1,1}\Big)\ \hat\otimes\ \Big(B\otimes\ketw{1,1}\Big)
  \ =\ \Big(A\otimes B\Big)\otimes\Big(\ketw{1,1}\, \hat\otimes\, \ketw{1,1}\Big)\nn
  &=&\Big(\bigoplus_jC_j\Big)\otimes\ketw{1,1}
   \ =\ \bigoplus_j\ketw{\hat C_j}
\label{ABC}
\eea
where $A\otimes B=\bigoplus_jC_j$ is the fusion of the representations $A$ and $B$ in the
logarithmic minimal model.
This extended fusion prescription is readily seen to be both associative and commutative.
It is also immediately confirmed that $\ketw{1,1}$ is the identity of the ensuing fusion algebra
\begin{equation}
 \ketw{1,1}\,\hat\otimes\, \ketw{\hat A}
  =\Big((1,1)\otimes \ketw{1,1}\Big)\,\hat\otimes\,\Big(A\otimes\ketw{1,1}\Big)
  =\Big((1,1)\otimes A\Big)\otimes\ketw{1,1}=\ketw{\hat A}
\end{equation}

We proceed by identifying the integrable boundary
conditions corresponding to the $2p$ irreducible representations $\ketw{r,s}$ and to
the $2p-2$ reducible yet indecomposable rank-2 representations 
$\ketw{\R_r^a}$ in the extended picture
\bea
 \ketw{1,s}&\!\!:=\!\!&(1,s)\otimes \ketw{1,1}\;=\;\bigoplus_{n=1}^\infty \,(2n-1)\,(2n-1,s)\nn
 \ketw{2,s}&\!\!:=\!\!&\half (2,s)\otimes \ketw{1,1}\;=\;\bigoplus_{n=1}^\infty \,2n\,(2n,s)\nn
 \ketw{\calR_{1}^a}
  &\!\!:=\!\!&\calR_{1}^a\otimes\ketw{1,1}\;=\;\bigoplus_{n=1}^\infty \,(2n-1)\,\calR_{2n-1}^a\nn
 \ketw{\calR_{2}^a}
  &\!\!:=\!\!&\half\calR_{2}^a\otimes\ketw{1,1}\;=\;\bigoplus_{n=1}^\infty \,2n\,\calR_{2n}^a
\label{ketw}
\eea
where $s\in\mathbb{Z}_{1,p}$ and $a\in\mathbb{Z}_{1,p-1}$.
These expansions all follow by setting $m=1$ in (\ref{prop11gen}).
Similarly, from (3.15), fusions such as $(2m-1,s)\otimes \ketw{1,1}$ and $(2m,s)\otimes \ketw{1,1}$ 
do not add independent fusion generators to the list (\ref{ketw}). 
We also note that
\bea
 \ketw{2,s}&=&(1,s)\otimes\ketw{2,1}=\ketw{1,s}\,\hat\otimes\,\ketw{2,1}\nn
 \ketw{\R_2^a}&=&\R_1^a\otimes\ketw{2,1}=\ketw{\R_1^a}\,\hat\otimes\,\ketw{2,1}
\label{2sw}
\eea
It follows from the expansions in (\ref{ketw}) that the embedding structures 
of the rank-2 indecomposable representations $\ketw{\R_1^a}$ and $\ketw{\R_2^{p-a}}$ 
can be described by
\psset{unit=.25cm}
\setlength{\unitlength}{.25cm}
\begin{equation}
 \mbox{
 \begin{picture}(20,10)(-15,0)
    \unitlength=1cm
  \thinlines
\put(-3.05,1){$\ketw{\R_1^a}:$}
\put(0.85,2){$\ketw{2,a}$}
\put(-1.05,1){$\ketw{1,p-a}$}
\put(2,1){$\ketw{1,p-a}$}
\put(0.85,0){$\ketw{2,a}$}
\put(1.05,1){$\longleftarrow$}
\put(1.65,1.5){$\nwarrow$}
\put(0.65,1.5){$\swarrow$}
\put(1.65,0.5){$\swarrow$}
\put(0.65,0.5){$\nwarrow$}
 \end{picture}
}
\hspace{5cm}
 \mbox{
 \begin{picture}(20,10)(-8,0)
    \unitlength=1cm
  \thinlines
\put(-2.8,1){$\ketw{\R_2^{p-a}}:$}
\put(0.55,2){$\ketw{1,p-a}$}
\put(-0.4,1){$\ketw{2,a}$}
\put(2,1){$\ketw{2,a}$}
\put(0.55,0){$\ketw{1,p-a}$}
\put(1.05,1){$\longleftarrow$}
\put(1.65,1.5){$\nwarrow$}
\put(0.65,1.5){$\swarrow$}
\put(1.65,0.5){$\swarrow$}
\put(0.65,0.5){$\nwarrow$}
 \end{picture}
}
\label{RvsR}
\end{equation} 
where the horizontal arrows indicate the off-diagonal action of the Virasoro mode $L_0$.
Similar embedding patterns also appeared in \cite{FGST0504,FGST0512,GR07}.

With the definitions (\ref{ketw}) and employing the properties (\ref{prop11gen}) and (\ref{2sw}),
we now establish that our fusion prescription (\ref{ABC}) applied to the fusion rules
of the logarithmic minimal model yields the fusion algebra of the extended picture (\ref{extFusion}).

To do this, we first consider the fusion
\begin{equation}
 \ketw{2,1}\,\hat\otimes\,\ketw{2,1}=\frac{1}{4}\Big((1,1)\oplus(3,1)\Big)\otimes\ketw{1,1} 
    =\frac{1}{4}(1+3)\ketw{1,1}=\ketw{1,1}
\end{equation}
and conclude that the horizontal fusion algebra $\langle \ketw{2,1}\rangle_{1,p}$ in the extended picture
is isomorphic to the well-known $A_2$ fusion algebra. 
This readily implies that
\begin{equation}
 \ketw{r,s}\,\hat\otimes\,\ketw{r',s'}
   =\ketw{r\cdot r',1}\,\hat\otimes\,\Big(\ketw{1,s}\,\hat\otimes\,\ketw{1,s'}\Big)
\label{rsrs}
\end{equation}
Likewise, we find that
\bea
 \ketw{r,s}\,\hat\otimes\,\ketw{\R_{r'}^a}
  &=&\ketw{r\cdot r',1}\,\hat\otimes\,\Big(\ketw{1,s}\,\hat\otimes\,\ketw{\R_1^{a'}}\Big)   \nn
 \ketw{\R_r^a}\,\hat\otimes\,\ketw{\R_{r'}^{a'}}
  &=&\ketw{r\cdot r',1}\,\hat\otimes\,\Big(\ketw{\R_1^a}\,\hat\otimes\,\ketw{\R_1^{a'}}\Big)
\label{rsR}
\eea

The vertical ($r$- and $r'$-independent) parts of (\ref{rsrs}) and (\ref{rsR}) are examined
using the compact expressions (\ref{VirFusCompact}). Only the last of the three types of fusion
undergo a simplification in the extended picture, namely
\bea
 \ketw{\R_1^a}\,\hat\otimes\,\ketw{\R_1^{a'}}&=&\Big(\R_1^a\otimes\R_1^{a'}\Big)\otimes\ketw{1,1}\nn
 &=&2\Big(\bigoplus_{\ell=|p-a-a'|+1,\ \!{\rm by}\ \!2}^{p-|a-a'|-1}\ketw{\R_1^\ell}\Big)
   \oplus4\Big(\bigoplus_{\ell=\epsilon(p+a+a'+1),\ \!{\rm by}\ \!2}^{|p-a-a'|-1}\ketw{\R_1^\ell}\Big)\nn
  &\oplus&2\Big(\bigoplus_{\ell=|a-a'|+1,\ \!{\rm by}\ \!2}^{p-|p-a-a'|-1}\ketw{\R_2^\ell}\Big)
    \oplus4\Big(\bigoplus_{\ell=\epsilon(a+a'+1),\ \!{\rm by}\ \!2}^{|a-a'|-1}\ketw{\R_2^\ell}\Big)
\eea
It is also recalled that $\R_2^a\otimes\ketw{1,1}=2\ketw{\R_2^a}$.
In conclusion, the fusion algebra in the extended picture reads 
\bea
 \ketw{r,s}\,\hat\otimes\,\ketw{r',s'}&=&\Big(\bigoplus_{j=|s-s'|+1,\ \!{\rm by}\ \!2}^{p-|p-s-s'|-1}
   \ketw{r\cdot r',j}\Big)
    \oplus\Big(\bigoplus_{\ell=\epsilon(s+s'-p-1),\ \!{\rm by}\ \!2}^{s+s'-p-1}
    \ketw{\R_{r\cdot r'}^\ell}\Big)\nn
 \ketw{r,s}\,\hat\otimes\,\ketw{\R_{r'}^a}&=&\Big(\bigoplus_{\ell=|s-a|+1,\ \!{\rm by}\ \!2}^{p-|p-s-a|-1}
   \ketw{\R_{r\cdot r'}^\ell}\Big)
   \oplus2\Big(\bigoplus_{\ell=\epsilon(s-a-1),\ \!{\rm by}\ \!2}^{s-a-1}\ketw{\R_{r\cdot r'}^\ell}\Big)\nn
  &\oplus&2\Big(\bigoplus_{\ell=\epsilon(s+a-p-1),\ \!{\rm by}\ \!2}^{s+a-p-1}
    \ketw{\R_{3-r\cdot r'}^\ell}\Big)\nn
 \ketw{\R_{r}^a}\,\hat\otimes\,\ketw{\R_{r'}^{a'}}
  &=&2\Big(\bigoplus_{\ell=|p-a-a'|+1,\ \!{\rm by}\ \!2}^{p-|a-a'|-1}\ketw{\R_{r\cdot r'}^\ell}\Big)
   \oplus4\Big(\bigoplus_{\ell=\epsilon(p+a+a'+1),\ \!{\rm by}\ \!2}^{|p-a-a'|-1}
     \ketw{\R_{r\cdot r'}^\ell}\Big)\nn
  &\oplus&2\Big(\bigoplus_{\ell=|a-a'|+1,\ \!{\rm by}\ \!2}^{p-|p-a-a'|-1}\ketw{\R_{3-r\cdot r'}^\ell}\Big)
    \oplus4\Big(\bigoplus_{\ell=\epsilon(a+a'+1),\ \!{\rm by}\ \!2}^{|a-a'|-1}\ketw{\R_{3-r\cdot r'}^\ell}\Big)
\label{1pextended} 
\eea
where the subscript $3-r\cdot r'$ arises from $(r\cdot r')\cdot2$.
We recall that $\ketw{\R_r^0}\equiv\ketw{r,p}$, $r,r'\in\mathbb{Z}_{1,2}$, $s,s'\in\mathbb{Z}_{1,p}$,
$a,a'\in\mathbb{Z}_{1,p-1}$ and that $\epsilon(n)$ and $r\cdot r'$ are defined in
(\ref{epsn}) and (\ref{rdotr}), respectively.
It follows straightforwardly from the fusion rules (\ref{1pextended}) that 
\begin{equation}
 \ketw{\R_r^a}\,\hat\otimes\,\ketw{\R_{r'}^{a'}}=2\Big(\ketw{r,p-a}\oplus\ketw{3-r,a}\Big)\,\hat\otimes\,
  \ketw{\R_{r'}^{a'}}
\end{equation}
thus verifying that our fusion algebra in the extended picture
(\ref{1pextended}) is equivalent to (\ref{extFusion}).

\section{Discussion}

We have considered the logarithmic minimal models ${\cal LM}(1,p)$ from a lattice perspective and looked for integrable structures which, in the continuum scaling limit, reflect the ${\cal W}$-extended
 symmetry. Although Virasoro and extended conformal symmetries cannot be seen directly on a finite lattice, since the finite system size manifestly breaks these symmetries, it is possible to see the {\em shadow} of the extended conformal symmetry in the structure of the ensuing closed finite fusion algebras.

Extending ideas originating with Cardy~\cite{Cardy1,Cardy2}, the fusion of representations of the conformal algebra can be implemented on the lattice by combining integrable boundary conditions associated with these fusions on the left and right edges of a strip. For rational theories, such as the minimal models, these integrable boundary conditions are constructed~\cite{BP01}, as solutions to the boundary Yang-Baxter equation, by fusing (in a lattice formulation of fusion) a finite number of columns to the boundary.
However, in the context of the logarithmic minimal models ${\cal LM}(1,p)$ in the extended picture, such a construction with a {\em finite} number of columns cannot work. This is clear since it is known~\cite{RP0707} that such constructions lead to finite direct sums of Virasoro representations whereas, from character formulas, it is clear that the ${\cal W}$-representations must in fact correspond to {\em infinite} sums of such represenations.

In this paper, we have constructed $4p-2$ integrable boundary conditions of the ${\cal LM}(1,p)$ models,  associated with the $4p-2$ distinct ${\cal W}$-representations, as limits of fusions on the boundary of the lattice.
In effect, this limiting process implies that the boundary condition is described by an infinite number of columns. On the one hand, this introduces a new class of boundary conditions allowing an arbitrarily large number of defects to propagate in the bulk compared to the fixed upper bound on the number of defects that emerges in the Virasoro picture. On the other hand, it also means that the
${\cal W}$-extended fusion on the lattice requires a non-trivial normalization to be well-defined in the limit of a large system. When normalized appropriately, we have shown that these new boundary conditions possess some simple stability properties that enable us to deduce the ${\cal W}$-extended fusion rules~\cite{GK9606,FK07,GR07} from the known fusion rules~\cite{GK9604,RP0707} in the Virasoro picture.
The closure of this fusion algebra on a finite number of representations in the extended picture
is remarkable confirmation of the consistency of the lattice approach and a clear demonstration that the logarithmic minimal models also provide lattice realizations of symplectic fermions and other logarithmic theories with extended conformal symmetry. Explicit Cayley tables for ${\cal LM}(1,2)$ and ${\cal LM}(1,3)$ are given in Figure~4.

\vskip.5cm
\section*{Acknowledgments}
\vskip.1cm
\noindent
This work is supported by the Australian Research Council and by the Belgian Interuniversity Attraction Poles Program P6/02, through the network NOSY (Nonlinear systems, stochastic processes and statistical mechanics). PAP and JR thank Ilya Tipunin for discussions. PR is a Research Associate of the Belgian National Fund for Scientific Research (FNRS).

\newpage
\begin{landscape}
\pagestyle{empty}
\appendix
\def\hcR{\hat\calR}

\begin{figure}[p]
\bigskip
\scriptsize
$$
\renewcommand{\arraystretch}{1.5}
\begin{array}{c||cc|cc|cc}
\hat\otimes&0&1&{-\frac{1}{8}}&\frac{3}{8}&\hat{\R}_0&\hat{\R}_1\\[4pt]
\hline \hline
\rule{0pt}{14pt}
0&0&1&-\frac{1}{8}&\frac{3}{8}&\hat{\R}_0&\hat{\R}_1\\[4pt]
1&1&0&\frac{3}{8}&-\frac{1}{8}&\hat{\R}_1&\hat{\R}_0\\[4pt]
\hline
\rule{0pt}{14pt}
-\frac{1}{8}&-\frac{1}{8}&\frac{3}{8}&\hat{\R}_0&\hat{\R}_1&2(-\frac{1}{8})+2(\frac{3}{8})&2(-\frac{1}{8})+
2(\frac{3}{8})\\[4pt]
\frac{3}{8}&\frac{3}{8}&-\frac{1}{8}&\hat{\R}_1&\hat{\R}_0&2(-\frac{1}{8})+2(\frac{3}{8})&2(-\frac{1}{8})+
2(\frac{3}{8})\\[4pt]
\hline
\rule{0pt}{14pt}
\hat{\R}_0&\hat{\R}_0&\hat{\R}_1&2(-\frac{1}{8})+2(\frac{3}{8})&2(-\frac{1}{8})+2(\frac{3}{8})
&2\hat{\R}_0+2\hat{\R}_1&2\hat{\R}_0+2\hat{\R}_1\\[4pt]
\hat{\R}_1&\hat{\R}_1&\hat{\R}_0&2(-\frac{1}{8})+2(\frac{3}{8})&2(-\frac{1}{8})+2(\frac{3}{8})
&2\hat{\R}_0+2\hat{\R}_1&2\hat{\R}_0+2\hat{\R}_1
\end{array}
\label{Cayley12}
$$
\bigskip
$$
\mbox{}\hspace{-.3in}\mbox{}
\renewcommand{\arraystretch}{1.5}
\begin{array}{c||cc|cc|cc|cc|cc}
\hat\otimes&0&\frac{7}{4}&-\frac{1}{4}&1&-\frac{1}{3}&\frac{5}{12}&\hcR_{1}^1&\hcR_{2}^1
&\hcR_{1}^2&\hcR_{2}^2\\[4pt]
\hline\hline
\rule{0pt}{14pt}
0&0&\frac{7}{4}&-\frac{1}{4}&1&-\frac{1}{3}&\frac{5}{12}&\hcR_{1}^1&\hcR_{2}^1&\hcR_{1}^2&\hcR_{2}^2\\[4pt]
\frac{7}{4}&\frac{7}{4}&0&1&-\frac{1}{4}&\frac{5}{12}&-\frac{1}{3}&\hcR_{2}^1&\hcR_{1}^1&\hcR_{2}^2&\hcR_{1}^2\\[4pt]
\hline
\rule{0pt}{14pt}
-\frac{1}{4}&-\frac{1}{4}&1&0\!+\!(-\frac{1}{3})&\frac{7}{4}\!+\!\frac{5}{12}&\hcR_{1}^1&\hcR_{2}^1&
2(-\frac{1}{3})\!+\!\hcR_{1}^2&2(\frac{5}{12})\!+\!\hcR_{2}^2&2(\frac{5}{12})\!+\!\hcR_{1}^1&
2(-\frac{1}{3})\!+\!\hcR_{2}^1\\[4pt]
1&1&-\frac{1}{4}&\frac{7}{4}\!+\!\frac{5}{12}&0\!+\!(-\frac{1}{3})&\hcR_{2}^1&\hcR_{1}^1&
2(\frac{5}{12})\!+\!\hcR_{2}^2&2(-\frac{1}{3})\!+\!\hcR_{1}^2&2(-\frac{1}{3})\!+\!\hcR_{2}^1&2(\frac{5}{12})\!+\!\hcR_{1}^1\\[4pt]
\hline
\rule{0pt}{14pt}
-\frac{1}{3}&-\frac{1}{3}&\frac{5}{12}&\hcR_{1}^1&\hcR_{2}^1&(-\frac{1}{3})\!+\!\hcR_{1}^2&\frac{5}{12}\!+\!\hcR_{2}^2&2(\frac{5}{12})\!+\!2\hcR_{1}^1&2(-\frac{1}{3})\!+\!2\hcR_{2}^1&2(-\frac{1}{3})\!+\!2\hcR_{2}^1&2(\frac{5}{12})\!+\!2\hcR_{1}^1\\[4pt]
\frac{5}{12}&\frac{5}{12}&-\frac{1}{3}&\hcR_{2}^1&\hcR_{1}^1&\frac{5}{12}\!+\!\hcR_{2}^2&(-\frac{1}{3})\!+\!\hcR_{1}^2&2(-\frac{1}{3})\!+\!2\hcR_{2}^1&2(\frac{5}{12})\!+\!2\hcR_{1}^1&2(\frac{5}{12})\!+\!2\hcR_{1}^1&2(-\frac{1}{3})\!+\!2\hcR_{2}^1\\[4pt]
\hline
\rule{0pt}{14pt}
\hcR_{1}^1&\hcR_{1}^1&\hcR_{2}^1&2(-\frac{1}{3})\!+\!\hcR_{1}^2&2(\frac{5}{12})\!+\!\hcR_{2}^2&2(\frac{5}{12})\!+\!2\hcR_{1}^1&2(-\frac{1}{3})\!+\!2\hcR_{2}^1&4(-\frac{1}{3})+2\hcR_{2}^1+2\hcR_{1}^2&4(\frac{5}{12})+2\hcR_{1}^1+2\hcR_{2}^2&4(\frac{5}{12})+2\hcR_{1}^1+2\hcR_{2}^2&4(-\frac{1}{3})+2\hcR_{2}^1+2\hcR_{1}^2\\[4pt]
\hcR_{2}^1&\hcR_{2}^1&\hcR_{1}^1&2(\frac{5}{12})\!+\!\hcR_{2}^2&2(-\frac{1}{3})\!+\!\hcR_{1}^2&2(-\frac{1}{3})\!+\!2\hcR_{2}^1&2(\frac{5}{12})\!+\!2\hcR_{1}^1&4(\frac{5}{12})+2\hcR_{1}^1+2\hcR_{2}^2&4(-\frac{1}{3})+2\hcR_{2}^1+2\hcR_{1}^2&4(-\frac{1}{3})+2\hcR_{2}^1+2\hcR_{1}^2&4(\frac{5}{12})+2\hcR_{1}^1+2\hcR_{2}^2\\[4pt]
\hline
\rule{0pt}{14pt}
\hcR_{1}^2&\hcR_{1}^2&\hcR_{2}^2&2(\frac{5}{12})\!+\!\hcR_{1}^1&2(-\frac{1}{3})\!+\!\hcR_{2}^1&2(-\frac{1}{3})\!+\!2\hcR_{2}^1&2(\frac{5}{12})\!+\!2\hcR_{1}^1&4(\frac{5}{12})+2\hcR_{1}^1+2\hcR_{2}^2&4(-\frac{1}{3})+2\hcR_{2}^1+2\hcR_{1}^2&4(-\frac{1}{3})+2\hcR_{2}^1+2\hcR_{1}^2&4(\frac{5}{12})+2\hcR_{1}^1+2\hcR_{2}^2\\[4pt]
\hcR_{2}^2&\hcR_{2}^2&\hcR_{1}^1&2(-\frac{1}{3})\!+\!\hcR_{2}^1&2(\frac{5}{12})\!+\!\hcR_{2}^2&2(\frac{5}{12})\!+\!2\hcR_{1}^1&2(-\frac{1}{3})\!+\!2\hcR_{2}^1&4(-\frac{1}{3})+2\hcR_{2}^1+2\hcR_{1}^2& 4(\frac{5}{12})+2\hcR_{1}^1+2\hcR_{2}^2&4(\frac{5}{12})+2\hcR_{1}^1+2\hcR_{2}^2&4(-\frac{1}{3})+2\hcR_{2}^1+2\hcR_{1}^2
\end{array}
$$
\caption{Cayley tables of the fusion algebras of  
${\cal LM}(1,2)$ and ${\cal LM}(1,3)$ respectively in the extended picture. 
The irreducible representations $\ketw{r,s}$ are denoted $\D_{r,s}$.
To further facilitate a comparison with \cite{GK9606}, we denote the
indecomposable rank-2 representations $\ketw{\R_r^1}$ by $\hat{\R}_{r-1}$ in the case of ${\cal LM}(1,2)$
and $\ketw{\R_r^a}$  by $\hat{\R}_{r}^a$ in the case of ${\cal LM}(1,3)$.}
\label{Cayley13}
\end{figure}
\end{landscape}


\end{document}